%% file: paper.tex
\documentclass[twoside,fleqn]{article}
\usepackage{espcrc2,epsf}

\usepackage[hyperindex]{hyperref}

\input rgmac.latex

\newcommand\MSbar{\ensuremath{\overline{\rm MS}}}
\renewcommand\MeV{\ensuremath{{\rm MeV}}}
\renewcommand\GeV{\ensuremath{{\rm GeV}}}
\newcommand\CPT{{$\chi$}PT}
\newcommand\Dslash{{\setbox0=\hbox{$D$}\hbox to 0pt{\hbox to \wd0{\hfil/\hfil}\hss}\box0}}
\newcommand\pslash{{\setbox0=\hbox{$p$}\hbox to 0pt{\hbox to \wd0{\hfil/\hfil}\hss}\box0}}

\newcommand\mbar{\hbox{$\overline{m}$}}

\def\gsim{\mathrel{\raise2pt\hbox to 8pt{\raise -5pt\hbox{$\sim$}\hss{$>$}}}}
\def\rsim{\mathrel{\raise2pt\hbox to 8pt{\raise -5pt\hbox{$\sim$}\hss{$>$}}}}
\def\lsim{\mathrel{\raise2pt\hbox to 8pt{\raise -5pt\hbox{$\sim$}\hss{$<$}}}}

\title{Advances in the Determination of Quark Masses}

\author{T. Bhattacharya
        and
        R. Gupta\address{%
          MS B285, T-8, Los Alamos National Laboratory, 
 	  Los Alamos, New Mexico 87545, USA.}%
        \thanks{Editors' note: T.~Bhattacharya was originally invited to give 
		this talk at LATTICE 97, but could not attend. He was replaced by R.~Gupta. 
	        They agreed to a joint writeup.}
}

\begin{document}

\begin{abstract}

Significant progress has been made in the determination of the light
quark masses, using both lattice QCD and sum rule methods, in the last 
year. We discuss the different methods and review the status of current 
results. Finally, we review the calculation of bottom and charm masses.

\end{abstract}

\maketitle

\section{Introduction}

One of the primary goals of phenomenology is to determine the
fundamental parameters of the standard model. Of these, the five quark
masses $m_u, m_d, m_s, m_c, m_b$ are amongst the least well known, and
cannot be measured in experiments. They have to be inferred from the
pattern of the observed hadron spectrum (chiral perturbation theory
(\CPT), HQET, and lattice QCD approaches) or from the study of 2-point
correlation functions (sum rules and lattice QCD approaches).  This
review is mainly an assessment of lattice QCD results, however we
shall also summarize the results obtained using \CPT\ 
and sum-rules. In particular we shall evaluate the extent to which
these methods agree with each other and discuss recent developments in
each.

\CPT\ relates pseudoscalar meson masses to \(m_u\), \(m_d\), and
\(m_s\). However, due to the presence of an overall unknown scale in
the chiral Lagrangian, \CPT\ can predict only two ratios amongst the
three light quark masses \cite{gasserPR,rMq90leutwyler,rMq96leutwyler}

\medskip\noindent
\begin{tabular}{ccc}
                          &  Lowest order & Next order \\
$ 2  m_s / (m_u + m_d) $  &  $24.2-25.9$  & $  24.4(1.5)  $ \phantom{\ .} \\ 
$           m_u / m_d  $  &  $0.55  $     & $  0.553(43)  $ . \\
\label{tab:mqcpt}
\end{tabular}

\smallskip
These ratios have been calculated neglecting the Kaplan-Manohar
symmetry \cite{KaplanManohar}. This is justified by assuming that the
higher order terms are small as discussed in \cite{rMq96leutwyler}.
These ratios, when combined with an absolute value determined from
sum-rules completed the ``standard'' scenario.  The status of sum-rule
estimates, as of 1996 was, $2\mbar \equiv m_u + m_d = 9.4(1.8) \MeV$
\cite{BPR95} and $m_s = 126(13) \MeV$ \cite{Jamin95}, evaluated in the
$\MSbar$ scheme at scale $\mu=2\GeV$.  These, combined with \CPT\ ratios, give
the standard estimates (in \MeV)

\medskip
\noindent\hspace{-6pt}
\begin{tabular}{lcccc}
Input    &  $m_u$     & $m_d$   & $\mbar$  & $m_s$    \\
$m_s$    &  3.7(1.0)  &6.7(0.8) &5.2(0.6) &   -      \\  
$\mbar$  &  3.3(1.3)  &6.1(1.1) & -       &  115(23) .\\
\label{tab:mscpt}
\end{tabular}
These values are consistent with observed spectra, $i.e.$ one can
explain the electromagnetic splittings, and the SU(3) breaking in the
mesons and in the baryon octet and decuplet without recourse to large
non-linear terms in $m_q$. The limitation of this combined analysis,
especially of the absolute numbers, is that there are no independent
checks as the quark masses enter in phenomenology only in combination
with other unknown quantities like the quark condensate. In the last
year there has been considerable activity in sum-rule analyses, and we
shall summarize these in Section~\ref{s:sumrules}.

Lattice QCD is a relative newcomer. Last year a major step forward was
taken in the understanding and quantification of systematic errors. An
analysis of the global data consolidated the lattice predictions of
significantly lower light quark masses \cite{Mq96LANL,Mq96GOUGH}
\begin{center}
\begin{tabular}{ccc}
                   &  Quenched    & $N_f=2$         \\
$ (m_u + m_d)/2 $  &  $3.4(7) $   & $\sim 2.7$ \MeV \\
$     m_s       $  &  $100(31)$   & $\sim 70 $ \MeV \,. \\
\label{tab:mq96lanl}
\end{tabular}
\vskip -0.5cm
\end{center}
The lattice results for the quenched theory
were considered reliable as three different discretizations of the
Dirac action, Wilson, clover, and staggered, gave results consistent
after extrapolation to the continuum limit.  The unquenched estimates
($N_f=2$ theory results) were preliminary. The main message was that
including two flavors of dynamical quarks lowers the quenched
estimates by $\sim 20\%$.

The abbreviation ``1996 data'' will be used to denote lattice data
analyzed in \cite{Mq96LANL}.  This will be used as the benchmark
against which progress will be measured.  Other reviews and references
to lattice results can be found in
\cite{Mq91Ukawa,Mq94Gupta,Mq96Mackenzie}.  The main sources of
uncertainty in the lattice estimates we shall explore are (i) the
extrapolation of the data to the continuum limit, (ii) the matching
constants between the lattice and continuum scheme,
and (iii) the effect of sea quarks (corrections due to quenching).  We
are happy to report that there has been significant progress in all
three areas in the last year.

\section{Lattice Approach: From Spectroscopy}

Lattice calculations need to determine four independent quantities to
predict the three light quark masses.  The fourth is needed to fix the
lattice spacing $a$. To distinguish between $m_u $ and $m_d$, it is
necessary to include electromagnetic corrections. This can be done by
including a U(1) field in the simulations \cite{Mq96Duncan}, however,
since the effect is small compared to statistical and systematic
errors, current lattice simulations have, for the most part, neglected
it.  Thus, this review is restricted to the determination of the
isospin symmetric combination $\mbar$.

A brief outline of how
the three quantities $\mbar$, $m_s$, and $a$ are determined from the
light hadron spectrum is as follows.
Using \CPT\ as the guiding principle, 
one writes down the most general expansion of hadron masses in terms of 
quark masses $m_1, m_2, m_3$, 
\begin{eqnarray}
M_\pi^2 a^2    \mskip-3\thinmuskip &=& \mskip-3\thinmuskip
	 A_\pi +    B_\pi a(m_1 + m_2)/2          + \ldots  \nonumber \\
M_\rho  a      \mskip-3\thinmuskip &=& \mskip-3\thinmuskip
	 A_\rho +   B_\rho a(m_1 + m_2)/2         + \ldots  \nonumber \\
M_{\Sigma^+} a \mskip-3\thinmuskip &=& \mskip-3\thinmuskip
	 A_{\Sigma^+}  + 4F am_1 + 2(F-D) am_3    + \ldots \nonumber \\
M_\Delta a     \mskip-3\thinmuskip &=& \mskip-3\thinmuskip
	 A_\Delta + B_\Delta a(m_1 + m_2 + m_3)/3 + \ldots  ,
\label{eq:chiralexp}
\end{eqnarray}
where for brevity we shall use $\pi, \rho, \Delta$ to denote
members of the pseudoscalar and vector octet and the
baryon decuplet respectively. Chiral symmetry implies 
$A_\pi \equiv 0$.   We leave it as a free parameter in the fits -- it gives a measure of
the uncertainty in the zero of the mass scale.

The simplest scenario is that there are no higher order corrections to
Eq.~\ref{eq:chiralexp}. Then, any triplet like $B_\pi, A_\rho, B_\rho$
(or equivalently $B_\pi, A_\Delta, B_\Delta$) can be used to
determine $\mbar$, $m_s$, and $a$.  For example, using the first set,
the quark masses are determined as 
\begin{equation}
   \mbar = \frac{M_\pi^2}{B_\pi a^{-1}} \, ; \, 
   m_s   = \frac{2(M_{K^*}-A_\rho a^{-1})}{B_\rho} - \mbar \,,
\end{equation}
and the scale from $M_\rho$ by solving the quadratic
\begin{equation}
A_\rho (\frac{1}{a})^2 - M_\rho (\frac{1}{a}) + B_\rho \frac{M_\pi^2}{B_\pi} = 0 \, .
\end{equation}
Similar relations exist for other choices of observables. 
The  scale $1/a$ can, in fact, be taken
from other observables like $M_n$, $M_\Delta$, $f_\pi$,
string tension, $r_0$, or the $1P-1S$ splitting in quarkonia. 
We choose $a(M_\rho)$ based on its ready availability and statistical
quality of $M_\rho$ versus $M_n$, $M_\Delta$ and $f_\pi$ data.
Different choices lead to different results, an unavoidable
uncertainty inherent in the quenched approximation. We shall refer to
this hadron spectroscopy method by the abbreviation HS. \looseness-1

If linearity were exact, or the exact expansions in
Eq.~\ref{eq:chiralexp} were known, then any single set, like the
pseudoscalar octet masses, could be used to fix both $\mbar a$ and
$m_s a$.  There is no {\it a priori} reason to assume that the higher
order chiral corrections are negligible. In fact, using $M_\pi/M_K$ or 
$M_\eta/M_K$ give different estimates for $m_s / \mbar$ in lowest order
\CPT. Also, present lattice data show non-linearities in the
pseudoscalar and vector meson data if a sufficiently large range of
quark masses is chosen \cite{rev96gottlieb,HM96LANL}.  However, these
non-linearities are small, and are very hard to detect in the typical
range $m_s/3 < m_q < 2 m_s$ used to extract quark masses.  Thus,
unless otherwise stated, we have used just linear fits in the chiral extrapolation.

The more serious problem facing quenched simulations is that this
range cannot be extended easily to much smaller quark masses due to the
presence of artifacts called quenched chiral logs. (For recent reviews
and references to this body of work see \cite{Mq94Gupta,CPT96Sharpe}.)
These terms, which are not present in the normal chiral expansion, are
singular in the limit $m_q \to 0$ and are a consequence of the fact
that, in the quenched approximation, the $\eta'$ propagator has a
single and a double pole.  The approach, therefore, has been to fit the
quenched data in this limited range, where these artifacts are
small, keeping just the normal chiral expansion. The
quenching error is then the change in these coefficients as sea quark
effects are added.

Finally, a test of the combined uncertainties due to quenching, and
chiral and $a\to 0$ extrapolations is that quark masses extracted
using different observables should all give the same results.  We
shall discuss the extent to which this is satisfied by comparing $m_s$
extracted from $M_K$ with that from $M_{K^*}$ or $M_\phi$ (labeled
$m_s(M_K)$, $m_s(M_{K^*})$, $m_s(M_\phi)$ respectively).  Most of the
figures we shall present will be for $\mbar$, however, note that due
to the linear approximation in Eq.~\ref{eq:chiralexp}, the result for
$m_s(M_K)$ is $ \equiv 25.9 \mbar$. $m_s(M_\phi)$ gives an independent
estimate, but the statistical signal in the data for vector state
masses is not as good as for pseudoscalars.  This will be evident from
the figures we show for $m_s(M_\phi)$.

\subsection{Definition of quark mass}

For staggered fermions, the fits in Eq.~\ref{eq:chiralexp} are made 
as a function of the bare quark mass input into the simulations. For 
Wilson-like fermions, we define light quark masses by
\begin{equation}
a m_{bare}  = \log(1 + (\frac{1}{2\kappa} - \frac{1}{2\kappa_c} ))
\end{equation}
where $\kappa_c$ is the critical value of the hopping parameter at
which the lattice pion mass vanishes. (Reference~\cite{Mq96LANL} used 
$a m_{bare} = 1/2\kappa - 1/{2\kappa_c}$, which is consistent to
$O(a)$. This change leads to differences of a few percent.)
So, compared to staggered fermions, Wilson like fermions
require the determination of an additional quantity $\kappa_c$.  The
need to calculate $\kappa_c$ can be avoided by using the Ward Identity
method described below.  From $m_{bare}$, the \MSbar\ mass is obtained
as
\begin{equation}
   m_{\MSbar} = Z_m m_{bare}
\end{equation}
where $Z_m = 1/Z_S$ is the matching factor between
$\MSbar$ and lattice schemes.  Since this factor turns out to be large
at 1-loop, its reliability has to be checked by non-perturbative
methods.

\subsection{Ward Identity (WI) method}

For Wilson like fermions, one can extract quark masses from
pseudoscalar mesons by using the axial vector ward identity
\begin{equation}
  Z_A \partial_\mu A_\mu(x)  = (m_1 + m_2) Z_P P(x) + \ldots
\end{equation}
where $A_\mu$ is the appropriate flavor non-degenerate bare axial
current $\bar \psi \gamma_\mu \gamma_5 \psi$, $P$ is the corresponding
pseudoscalar density $\bar \psi \gamma_5 \psi$, and the $Z$'s are the
corresponding renormalization constants. The dots represent
discretization corrections, whose size depends on the order to which
improvement of the action and operators has been carried out.  The
renormalized quark masses are then given by the ratio of 2-point
functions
\begin{equation}
  (m_1 + m_2) =   \frac{Z_A}{Z_P} 
      \frac{\langle \partial_\mu A_\mu(x) J(0) \rangle}{\langle P(x)  J(0) \rangle}  + \ldots
\end{equation}
where $J$ is any operator that couples to pions.  The advantage of
this $WI$ method is that $\kappa_c$ does not enter into the
calculation.  The limitation is that only the pseudoscalar sector is
tested.

\section{Details of the analysis of World Data}
\label{s:Adetails}

Different groups have their own favorite ways of doing the fits, and
of dealing with the two major sources of systematic errors, the
extrapolation to the continuum limit and the determination of
$Z$'s. Consequently, the same data analyzed by different groups can
lead to slightly different estimates of quark masses.  Since we wish
to do a global analysis, we have attempted to minimize these relative
differences. We start with the data for pseudoscalar and vector mesons
as a function of quark masses and $g^2$, and redo the analysis to
extract $A_\pi, B_\pi, A_\rho, B_\rho$.  For the physical masses we
use $M_\pi=135$ MeV, $M_\rho=770$ MeV, $M_K=495$ MeV, $M_{K^*}=894$
MeV, and $M_\phi=1020$ MeV.  (Ref.~\cite{Mq96LANL} used $M_\pi=137$,
thus $\mbar$ quoted there is about 3\% higher.)  This still
leaves two sources of uncertainty. (i) The difference in lattice sizes
and the subjective bias in fits made to extract hadron masses. This we
check by comparing data from different collaborations. (ii) The
statistical correlations in the data. These require knowledge of the
covariance matrix which is not available in most cases.

To convert the lattice mass to a continuum scheme like \MSbar, we
require the calculation of either $Z_S$ in both lattice and continuum
renormalization scheme (HS method), or of $Z_A$ and $Z_P$ (WI method).
For these matching factors 1-loop results have been used in the past.
Now, non-perturbative estimates are becoming common. The calculation
of the Lepage-Mackenzie $\alpha_s$, ``horizontal'' matching between
the lattice and continuum schemes in the perturbative approach, details
of the error analysis, and the evolution in the continuum are the same
as in \cite{Mq96LANL}.  The relevant formula for the 2-loop evolution
needed with the 1-loop matching is \cite{Allton94}
\begin{equation}
{m(Q) \over m(\mu)}  = \bigg({g^2(Q) \over g^2(\mu)}\bigg)^{\gamma_0 / 2 \beta_0} \
        \bigg( 1 + {g^2(Q) - g^2(\mu) \over 16\pi^2} J \bigg) \,,
\label{eq:evolution}
\end{equation}
where $J = (\gamma_1 \beta_0 - \gamma_0 \beta_1)/ 2 \beta_0^2$. 
From this the renormalization group invariant mass $\widehat m$ is
defined as \looseness-1
\begin{equation}
\widehat m = m(\mu)  \bigg( {2\beta_0 g^2(\mu) \over 16\pi^2} \bigg)^{-\gamma_0 / 2 \beta_0} \
        \bigg( 1 - {g^2(\mu) \over 16\pi^2} J \bigg) .
\label{eq:RGImass}
\end{equation}
This evolution in the continuum or the conversion to $\widehat m$
introduce no new lattice uncertainty.

Now we briefly highlight two important developments by the APE and
ALPHA collaborations that overcome the uncertainty introduced by the
perturbative matching.

The APE collaboration has carried through non-perturbative
determination of the lattice $Z$'s \cite{Mq97Giusti} using the chiral Ward
identity method \cite{WI85ROME,WI95ROME}.  At present non-perturbative
estimates of all the necessary lattice $Z$'s for all the actions we
discuss are not known.  As these become available, the uncertainty in
using the 1-loop relations will be removed. We shall discuss the
impact of the known $Z$'s on current data.

The ALPHA Collaboration has initiated a very successfully
non-perturbative program to improve the discretization of the Dirac
action, fermionic operators, and their renormalization constants
\cite{Mq97ALPHA}.  The improvement they propose over APE's chiral Ward
Identity method for determining the $Z's$ is to eliminate making any
connection with a continuum scheme by directly computing the
renormalization group independent quark mass. The essential steps
in their method are (i) calculate the renormalized quark mass
using the WI method at some infrared lattice scale ($Z_P$ and $Z_A$
are also calculated non-perturbatively in the Schr\"odinger functional
scheme), (ii) evolve this result to some very high scale using the
step scaling function calculated non-perturbatively, and (iii) at this
high scale, where $\alpha_s$ is small, make contact with lattice
perturbation theory to define $\widehat m$.  This non-perturbatively
calculated $\widehat m$ is scheme independent, thus subsequent
conversion to a running mass in some scheme and at some desired scale
can be done in the continuum using the most accurate versions of the
scale evolution equations \cite{97Vermaseren}. No data for $\widehat
m$ has yet been released. The status of this approach has been
reviewed by L\"uscher at this conference \cite{Mq97ALPHA}.

\section{Quenched Wilson fermion results}

The Wilson formulation of the Dirac action is the simplest and has
been the most commonly used in numerical simulations until very
recently. Its disadvantage is that discretization errors and
violations of chiral symmetry begin at $O(a)$. Nevertheless, a 1996
compilation of the world data for quark masses \cite{Mq96LANL} showed
that an extrapolation to the continuum limit can be made keeping only
the lowest order corrections, giving 
\begin{eqnarray}
  \mbar       &=& 3.4(7)\MeV [1 + 1.3(2) \GeV\ a] \nonumber \\
  m_s(M_\phi) &=& 94(27)\MeV [1 + 1.9(7) \GeV\ a] .
\label{eq:wilson1996}
\end{eqnarray}
The surprise was the size of the $O(a)$ errors. The typical size of
the slope in $a$ obtained in the extrapolation of other observables
like $M_\rho, f_\pi, \ldots$ is a few hundred MeV, so the $1.2-1.7
\GeV$ values needed to be understood.

\begin{figure}[t] 
\vspace{9pt}
\hbox{\hskip15bp\epsfxsize=0.9\hsize \epsfbox {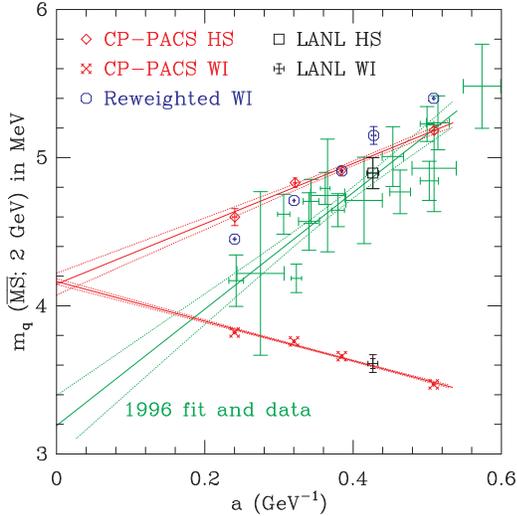}}
\vskip -0.8cm
\caption{Linear extrapolation of $\mbar$ versus $a(M_\rho)$ for Wilson fermions 
using HS and WI methods. The 1996 data and fit are shown in green.}
\vskip-0.60cm
\label{f:Wmbar}
\end{figure}

Significant progress in this sector has been made by the recent high
statistics, large lattice calculations by CP-PACS which have been 
reviewed by T.~Yoshie in 
\cite{LAT97Yoshie}. They have provided four new data points on
lattices of size $\geq 3$ fermi.  The new data are shown in
Fig.~\ref{f:Wmbar}.  To highlight the statistical improvement we show
data at $\beta=6.0$ from the next best calculation (with respect to
both statistics and lattice size) \cite{HM96LANL}.  The data from both
HS and WI methods are well fit by just a linear correction, and
extrapolate to roughly the same value
\begin{eqnarray}
   \mbar &=& 4.1(1) \MeV [ 1 + 0.49 \GeV\ a ] \qquad {\rm HS} \nonumber \\
   \mbar &=& 4.2(1) \MeV [ 1 - 0.32 \GeV\ a ] \qquad {\rm WI}.
\end{eqnarray}
The CP-PACS data, therefore, gives a significantly smaller slope and 
a larger value for $\mbar$. 

In Fig.~\ref{f:Wmbar} we also show the 1996 fit. The change in the
slope is due to a pivot about $\beta \approx 5.9$. Further analysis
based on the CP-PACS data shows that the three APE points at
$\beta=6.3, 6.4$, that biased the 1996 fits towards lower values of
$\mbar$, are not compatible within errors with CP-PACS; the estimates
of both $B_\pi$ and $a$ differ. Our conclusion is that these data
should not be used in ``modern analyses''.\looseness-1

\begin{figure}[t]  
\vspace{9pt}
\hbox{\hskip15bp\epsfxsize=0.9\hsize \epsfbox {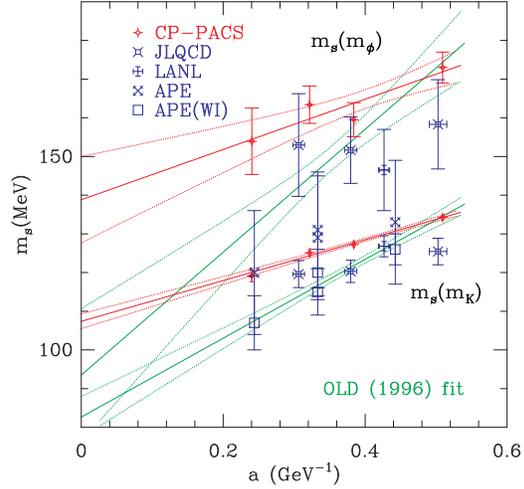}}
\vskip -0.8cm
\caption{Linear extrapolation of $m_s(M_K)$ and $m_s(M_\phi)$
versus $a(M_\rho)$ for Wilson fermions. 
The 1996 fits are shown in green for comparison.}
\vskip-0.60cm
\label{f:Wms}
\end{figure}

The CP-PACS collaboration also calculate $m_s$ using both $M_K$ and
$M_\phi$ ($M_\phi$ and $M_{K^*}$ give consistent results). The new data 
are shown in Fig.~\ref{f:Wms}. The data show 
that $m_s(m_K)$ and $m_s(M_\phi)$ are different 
even after extrapolation to the continuum limit, as was 
the case in the 1996 data, 
\begin{eqnarray}
 m_s\,(M_\phi) &=& 139(11) \MeV [ 1 + 0.47 \GeV\ a ]\nonumber\\
   m_s (M_K)   &=& 107(2) \MeV [ 1 + 0.49 \GeV\ a ] .
\end{eqnarray}
Assuming that this difference is not an artifact of statistical errors
or due to the uncertainty in the extrapolation, it could be due either
to quenching, or to the failure of the assumption of linearity in the
chiral fits. In that case one will need unquenched data to
resolve this issue. Meanwhile, we consider this difference as one of
the remaining systematic errors.  These estimates for $m_s$ are larger
than those reported in 1996. The reasons for the change are the same
as for $\mbar$.

The second important contribution of the CP-PACS analysis is the
reconciliation of the HS and WI methods.  The data, and the fits shown
in Fig.~\ref{f:Wmbar} indicate that the difference between the two
methods at finite $a$ is due to different $O(a)$ discretization
errors.  Giusti \etal~\cite{Mq97Giusti} argue that this difference is
due to the failure of the 1-loop $Z$'s.  We have therefore reanalyzed
the CP-PACS/LANL WI data using the non-perturbative $Z_A, Z_P$
calculated by Giusti \etal\ (we interpolated or extrapolated the
Giusti \etal\ results to other $\beta$ linearly).  These reweighted
points are plotted using the symbol octagon in Fig.~\ref{f:Wmbar}.  What is remarkable is
the agreement between non-perturbative WI and perturbative HS data
around $\beta = 6.1$, and the change in the slope, $i.e.$ the slope
switches sign and ends up being even larger than that for HS with
1-loop $Z$'s!  This equality of HS and WI data around $\beta = 6.1$
is consistent with the findings of Giusti \etal~\cite{Mq97Giusti} as
discussed in section \ref{s:APE}.  What the APE data cannot expose is
the slope in $a$ from simulations at just $\beta=6.0$ and $6.2$.  We
would also like to point out that $Z^{non-pert} / Z^{pert}$
typically decreases rapidly with $\beta$ as, for example, in
Table~\ref{t:APE}. Hence, if the 1-loop $Z_m$ turns out to be 
an underestimate, then correcting for it will increase the slope of the
HS data as well.  Thus, the issue of
the size of the slope, and of the final extrapolated value will be
resolved soon, once the non-perturbative estimates of $Z_m$ are made
available \cite{Mq97Giusti}.

\subsection{Clover Action}

\begin{figure}[t] 
\vspace{9pt}
\hbox{\hskip15bp\epsfxsize=0.9\hsize \epsfbox {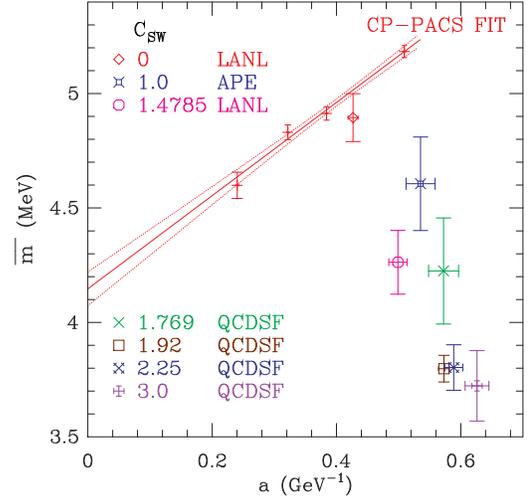}}
\vskip -0.8cm
\caption{Behavior of $\mbar$ and $a(M_\rho)$ as a function of the
clover coefficient $C_{SW}$ at $\beta=6.0$. Also shown 
are the CP-PACS data and fit from Fig.~\protect\ref{f:Wmbar}.}
\vskip-0.60cm
\label{f:SWimp}
\end{figure}

One way to reduce/remove the large $O(a)$ corrections in the Wilson
formulation, and thus improve the reliability of the $a=0$
extrapolation, is to simulate the Sheikholeslami-Wohlert (SW or
clover) action.  In the last couple of years a number of calculations
have been done using this action. Unfortunately, different
calculations use different values of $C_{SW}$: (i) tree-level
value $C_{SW}=1$, (ii) tree-level Tadpole Improved (TI), (iii) 1-loop
tadpole improved, and (iv) non-perturbative $O(a)$ improved. To get a
feel for the effects of tuning $C_{SW}$, we first show in
Fig.~\ref{f:SWimp} the data for $\mbar$ for different values of
$C_{SW}$ as a function of $a$.  The data at $\beta=6.0$ show that as
$C_{SW}$ is increased, $\mbar$ decreases and $a$, as determined from
$M_\rho$, increases.  The expectation is that as $\beta \to \infty$,
the ordering should stay the same, only the spread should decrease.

\begin{figure}[t] 
\vspace{9pt}
\hbox{\hskip15bp\epsfxsize=0.9\hsize \epsfbox {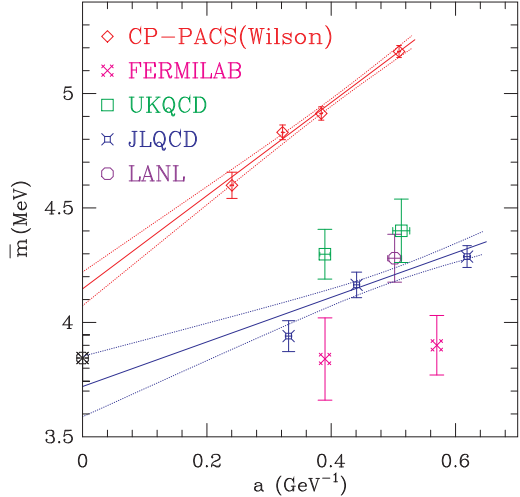}}
\vskip -0.8cm
\caption{$\mbar$ from the TI clover data versus $a(M_\rho)$. 
Also shown are CP-PACS Wilson data and fit from Fig.~\protect\ref{f:Wmbar}. Result of 
extrapolation in $\alpha a$ is shown by the fancy square at $a=0$.}
\vskip-0.60cm
\label{f:SWTmbar}
\end{figure}

\begin{figure}[t]  
\vspace{9pt}
\hbox{\hskip15bp\epsfxsize=0.9\hsize \epsfbox {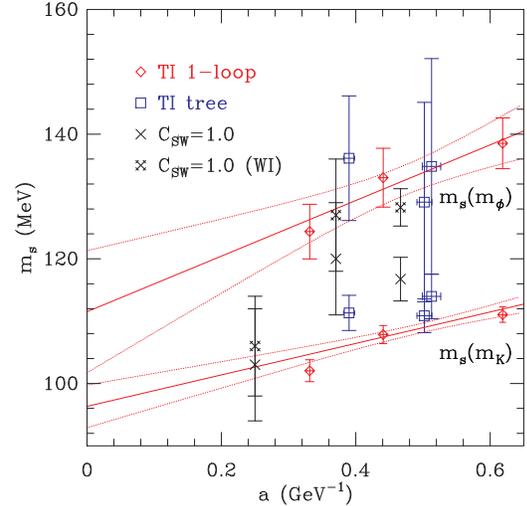}}
\vskip -0.8cm
\caption{Data for $m_s(M_K)$ and $m_s(M_\phi)$
versus $a(M_\rho)$ for TI clover fermions, along with simple 
linear extrapolation. 
Also shown are APE data with $C_{SW}=1$ for comparison.}
\label{f:SWTms}
\vskip-0.60cm
\end{figure}

The most extensive data are using the tadpole improved clover action
(we do not distinguish between tree-level and 1-loop improved as the
difference in $C_{SW}$ is small and we assume that it amounts to a
negligible change in quark masses). These data for $\mbar$ from JLQCD
\cite{Mq97JLQCD,Mq97WIJLQCD}, LANL \cite{Mq97LANL}, and UKQCD
\cite{Mq97UKQCD} collaborations are shown in Fig.~\ref{f:SWTmbar}.  We
also show the Fermilab data given in \cite{Mq96GOUGH}.  Of the
$\approx 10\%$ difference between JLQCD and Fermilab data at
$\beta=5.9$ and $6.1$, $\approx 6\%$ comes from the setting of $a$
(Fermilab uses $1P-1S$ splitting in charmonium), and the rest from
differences in $B_\pi$ and conversion to $\MSbar$.  Since the data
needed for extracting $a(M_\rho)$ for the Fermilab runs are not
available, we cannot use these points in our continuum
extrapolation. A fit to the rest of the data, assuming that the errors
are $O(\alpha a)$, gives \looseness -1
\begin{equation}
  \mbar = 3.84(10) \MeV [ 1 + 1.0 \GeV\ \alpha(a)\ a ], 
\label{eq:SWTmbar}	
\end{equation}
shown by the fancy square at $a=0$. An analysis of $m_s(M_K)$ and $m_s(M_{\phi})$ gives 
\begin{eqnarray}
   m_s(M_\phi) \mskip-\thinmuskip &=& \mskip-\thinmuskip
		117(8) \MeV [ 1 + 1.5 \GeV \alpha(a) a ]\nonumber\\ 
   m_s (M_K)   \mskip-\thinmuskip &=& \mskip-\thinmuskip
		99(3) \MeV [ 1 + 1.0 \GeV \alpha(a) a ] .
\end{eqnarray}
In the figure we plot, for simplicity, the linear fit which has roughly the 
same $\chi^2$
\begin{equation}
  \mbar = 3.72(13) \MeV [ 1 + 0.26 \GeV\  a ].
\end{equation}
This lies $\approx 1 \sigma$ below our preferred fit and value given in
Eq.~\ref{eq:SWTmbar}. Similar linear fits to the $m_s(M_K)$ and
$m_s(M_{\phi})$ data are shown in Fig.~\ref{f:SWTms}.

%

\begin{figure}[t]  
\vspace{9pt}
\hbox{\hskip15bp\epsfxsize=0.9\hsize \epsfbox {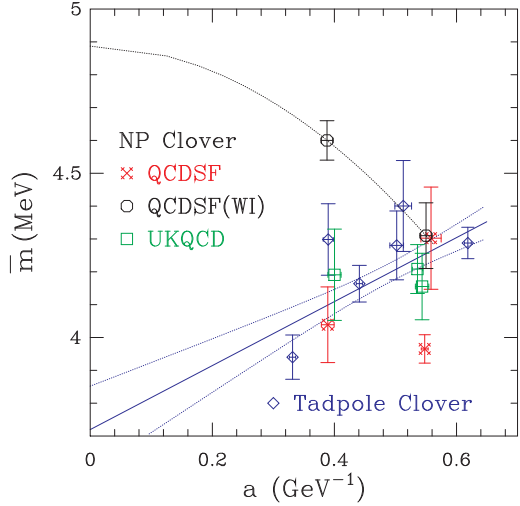}}
\vskip -0.8cm
\caption{$\mbar$ from non-perturbative SW fermions. 
TI clover fit from Fig.~\ref{f:SWTmbar} is also shown.}
\label{f:NPSWmbar}
\vskip-0.60cm
\end{figure}

The APETO \cite{Mq97APETO}, QCDSF \cite{Mq97QCDSF} and UKQCD
\cite{Mq97UKQCD} collaborations have calculated quark masses using the
non-perturbative value of \(C_{SW}\) determined by the ALPHA
collaboration.  The data for $\mbar$ by QCDSF and UKQCD at $\beta=6.0$
and $6.2$ are shown in Fig.~\ref{f:NPSWmbar} and compared to tadpole
improved clover.  We find the following features. At $\beta=6.0$, data
with the WI (QCDSF) and HS (larger volume QCDSF and UKQCD) methods
agree, and roughly lie on the tadpole improved clover line. Going to $
\beta=6.2$, the QCDSF WI data show a small increase.  Consequently, a
fit to the two QCDSF-WI points, assuming $O(a^2)$ errors, gives the
large value $\mbar = 5.1(2)$ MeV (the value in the
Fig.~\ref{f:NPSWmbar} is slightly different as it is from our
analysis). On the other hand the QCDSF-HS value decreases slightly,
while the UKQCD-HS data (still preliminary) show no $a$ dependence.
Our overall conclusion is that the range in $\beta$ is too narrow to
allow a meaningful extrapolation in $a$ with just two points.  

Both APETO \cite{Mq97APETO} and QCDSF \cite{Mq97QCDSF} have extracted
$m_s$ at $\beta=6.2$ on $24^3 \times 48$ lattices. Their raw data for
pseudoscalar and vector masses are consistent.  Since the extraction
of $m_s$ includes $ma$ corrections to the renormalization constants,
their results are correct to $O(a)$.  The estimates of $m_s$ are
$111(15)$ and $102(2)$ MeV respectively. QCDSF find a small increase
between $\beta=6.0$ and $\beta=6.2$, and their value extrapolated to
$a=0$ is $112(5) $ MeV.  Higher statistics and larger lattices data at
other values of $\beta$ are expected in the next year. These will
improve the reliability of the extrapolation to the continuum limit.

\subsection{Staggered Fermions}

The advantage of staggered fermion formulation is the remnant chiral
symmetry that guarantees Ward identities as in the continuum.  As a
result $Z_A = 1$ and $Z_S = Z_P$, and the quark mass is only
multiplicatively renormalized. Consequently, the two methods for
determining quark masses, HS and WI, are the same. The sticky point is
that the finite piece in TI 1-loop expression $Z_m = 1/Z_P^{local} =
1 - \frac{\alpha_s}{4\pi} ( 8 \log(\mu a ) - 39.1)$ is very large. Therefore, the
1-loop matching between lattice and continuum may not be reliable.
Second, the 1996 data at $\beta > 6.3 $ was preliminary and suggested
a small increase with $\beta$.  There has been progress on both fronts
in the last year.  First, a partially non-perturbative analysis of the
reliability of $Z_P$ has been done by Gupta, Kilcup, and Sharpe
\cite{stagrenorm}, and secondly there is new data by the 
JLQCD \cite{MqS97JLQCD} and MILC \cite{MqS97MILC} collaborations.

The basic idea of the partially non-perturbative analysis is that if 
$Z_P^{local}$ is large, then one could get a more reliable estimate using 
\begin{equation}
Z_P^{local} = Z_P^{smeared} \times \left( {Z_P^{local} \over Z_P^{smeared} } \right)
\end{equation}
where ``smeared'' is any discretization of the pseudoscalar density
that has a reliable perturbative expansion. The ratio in the
parenthesis, which is a large factor, is calculated
non-perturbatively. On the basis of such an analysis, Gupta, Kilcup,
and Sharpe found that the 1-loop perturbative expression for
$Z_P^{local}$ is $\sim 5\%$ smaller than the partially
non-perturbative result at $\beta=6.0$.  The caveat in this
calculation is that the non-perturbative determination of the ratio
shows large $O(a^2)$ discretization errors, and the extrapolation to
$a = 0$ is based on only two points at $\beta=6.0$ and $6.2$. Thus,
the calculation of the ratio needs corroboration.  For the moment we
apply this shift when presenting final estimates in
Section~\ref{s:qfinal}.\looseness-1

\begin{figure}[t]  
\vspace{9pt}
\hbox{\hskip15bp\epsfxsize=0.9\hsize \epsfbox {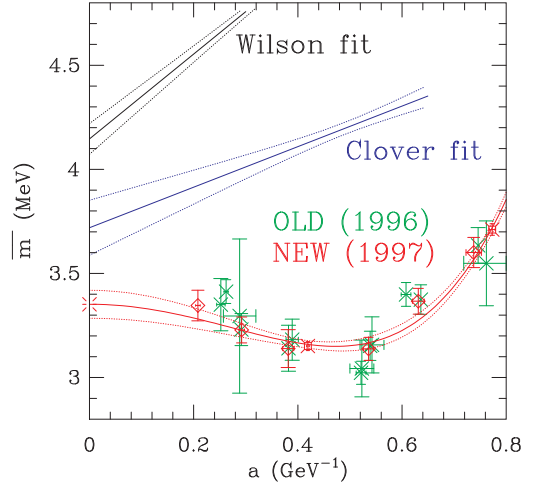}}
\vskip -0.8cm
\caption{Staggered fermion data for \mbar\ versus $a(m_\rho)$. The fit
assumes $a^2$ and $a^4$ corrections.}
\label{f:mbarstag}
\vskip-0.60cm
\end{figure}

The staggered fermion world data are shown in
Fig.~\ref{f:mbarstag}. The new data by JLQCD and MILC collaborations
are consistent with the old, and confirm the small rise at weaker
couplings.  Assuming that the partially non-perturbative analysis of
$Z_P$ validates the 1-loop result, and thus the non-monotonic
behavior, we fit the data including $a^2$ and $a^4$ corrections. (Note
that the absence of $O(a)$ corrections implies that the fit approaches
$a=0$ with zero slope.) The results, in \MeV,  are
\begin{eqnarray}
\mbar\,     \mskip-4\thinmuskip &=& \mskip-4\thinmuskip
        3.35(7) \big[ 1 {- (.7\GeV\,a)^2} {+ (1\GeV\,a)^4} \big] \nonumber \\
m_s(M_\phi) \mskip-4\thinmuskip &=& \mskip-4\thinmuskip
	104(5)  \big[ 1 {- (1\GeV\,a)^2} {+ (1\GeV\,a)^4} \big] .
\end{eqnarray}
The discretization corrections are small, and account for the
non-monotonic behavior, however, to resolve this issue a fully
non-perturbative calculation of $Z_m$ is needed.  The net upshot is
that the staggered results using 1-loop $Z_m$ are basically
unchanged, and raised by $\sim 5\%$ by using the partially
non-perturbative calculation of $Z_m$.

\subsection{Non-perturbative $Z$'s: APE data}
\label{s:APE}

The APE collaboration has presented new data on the estimates of
$\mbar, m_s$, and $m_c$ using both the HS and WI methods
\cite{Mq97Giusti}.  They also provide a non-perturbative determination
of $Z_A, Z_P$ needed in the WI method, however, the $Z_m$ used in
the HS method is still 1-loop.  In their calculations of the NP $Z$'s
for Wilson fermions bare operators are used, while for clover fermions
a field rotation by $1 + a \Dslash/4$ is included.

The APE data are summarized in Table~\ref{t:APE}.  We have also included a
comparison of the tadpole improved $ Z_A/Z_P$ with the
non-perturbative factor $ (1 + \alpha_s C_m^{\rm Lan}) Z_A / Z_P^{RI}$
with $\mu^2 a^2 = 0.9648$.  In keeping with the authors, who consider
the $\beta=6.4$ lattices too small, we neglect the corresponding quark
mass data. The first remarkable feature is that Wilson and clover
fermions give consistent results for the WI method.  Next, based on
the consistency of the data at $\beta=6.0$ and $6.2$, the authors
assume that there are no significant discretization errors.  Thus,
their final number $m_s(M_K) = 123\pm 4 \pm 15$ is from the WI method,
averaged over the Wilson and Clover fermions data at $\beta=6.0$ and
$6.2$.

We consider the agreement between WI and the HS data fortuitous.  If
the difference between the perturbative and non-perturbative $Z_m$ is
similar to that in $Z_P$, then the final HS results would be significantly
different.  Second, the CP-PACS/LANL data show a significant $a$
dependence which the APE data are not precise nor extensive enough to
expose. This is the reason why their final value is higher than the
world average given next.

\begin{table} 
\caption{APE data for $m_s(M_K)$ for Wilson (upper box) and $C_{SW}=1$
clover fermions. The perturbative and non-perturbative $Z$'s are also
compared.}
\setlength\tabcolsep{5pt}
\input {t-ape}

\label{t:APE}
\vskip-0.60cm
\end{table}

\section{Bottom line on quenched results}
\label{s:qfinal}

A summary of the quenched results, in \MeV, is 

%
%
\medskip\noindent
\begin{tabular}{lccc}
                 &  Wilson    & TI Clover   &  Staggered    \\
$ \mbar       $  &  $4.1(1)$  & $3.8(1)$    & $ 3.5(1)$     \\
$ m_s(M_K)    $  &  $107(2)$  & $ 99(3)$    & $ 91(2)$      \\
$ m_s(M_\phi) $  &  $139(11)$ & $117(8)$    & $ 109(5)$  .  \\
\label{tab:mqfinal}
\end{tabular}

\smallskip

These differences between Wilson, TI clover, and staggered results
could easily be accounted for by the uncertainties in the
extrapolations, and/or by the ratio of the non-perturbative to 1-loop
estimates of $Z$'s. Thus, for our best estimate we take the mean and
use the spread as the error.  In case of $m_s$ we have an addition
uncertainty coming from the dependence on the state used to extract
it, $M_K$ versus $M_{K^*}$ (or equivalently $M_\phi$).  This could be
due to the quenched approximation or an artifact of having used linear
chiral extrapolations. Since we do not have control over either of
these two uncertainties, we again average over all the data.  To
these, we add a second uncertainty of $10\%$ as due to the
determination of the scale $1/a$. Thus, our best estimates of $\MSbar$
masses evaluated at $2 $ GeV are
\begin{eqnarray}
\mbar &=& 3.8(4)(4)   \ \MeV  \nonumber\\
m_s   &=& 110(20)(11) \ \MeV  \,.
\label{eq:mqfinal}
\end{eqnarray}

\section{$N_f = 2$ Wilson results}

The SESAM collaboration has presented new data for $n_f=2$ Wilson
fermions at $\beta=5.6$ \cite{Mq97SESAM}.  The novel feature of their
data is that pseudoscalar and vector masses have been calculated for
degenerate ($\kappa^V = \kappa^{sea}$), and for non-degenerate
($\kappa^V \ne \kappa^{sea}$) combinations. The superscripts $V$ and
${sea}$ refer to valence and sea quarks respectively.  Their data 
can be fit to 
\begin{eqnarray}
  M_\pi^2 &=&  A_\pi + B_\pi^V \frac{1}{2 \kappa^V} + B_\pi^{sea} \frac{1}{2 \kappa^{sea}} \nonumber \\
  M_\rho  &=&  A_\rho + B_\rho^V \frac{1}{2 \kappa^V} + B_\rho^{sea} \frac{1}{2 \kappa^{sea}} \,.
\end{eqnarray}
As a result they study the dependence on both $\kappa^V$ and $\kappa^{sea}$ .  

The first surprise of their calculation is a large dependence on
\(\kappa^{sea}\) (\(B^V \sim B^S\)).  This is surprising because QCD
perturbation theory and the success of lowest order \CPT\ suggests
that the dependence on sea quark masses should be small.

Their second important contribution is that they correctly point out that
all previous extractions of $m_s$ from $n_f=2$ data are incorrect. 
Previous calculations were essentially calculating $m_s$ in a sea 
of strange quarks. 

The last issue they raise is specific to Wilson-like fermions and
concerns the zero of mass scale ($\kappa_c$), and concomitantly the
definition of the quark mass.  They discuss two possible ways of
analyzing the data. (i) Degenerate extrapolation: determine
$\kappa_c^0$ by extrapolating pion masses for the degenerate case
$\kappa^V = \kappa^{sea}$ and measure all quark masses with respect to
this $\kappa_c^0$.  (ii) Partially quenched extrapolation: calculate
$\mbar$, $m_s$ and $a$ for a fixed sea quark mass, and then
extrapolate in the sea quark mass. They argue that the first method is
the correct one and discard the second.  It is at this point we
disagree with them.  What we now show is that if one uses the
conventional perturbative $Z_m$ to relate the lattice mass to $\MSbar$
scheme, then it is actually the partially quenched method that is more
appropriate. We also argue that if $Z_m$ is calculated
non-perturbatively, then both methods would give the same result for
$\mbar$. Lastly, our analysis shows that the large dependence on
\(\kappa^{sea}\) is actually that of $\kappa_c$ on \(\kappa^{sea}\),
and not of physical masses.

The pole mass defined by the inverse propagator
$ S_\psi^{-1} =  Z_\psi^{-1} ( \pslash - m ) + \cdots $, 
can be written as 
\begin{equation}
m a = \frac{1}{2\kappa^V} - 8 + 
         \delta m \left( \alpha_s, \frac{1}{2\kappa^V}, \frac{1}{2\kappa^{sea}} \right) \, .
\end{equation}
The linear divergence in $\delta m$ is absorbed in the 
definition of \(\kappa_c^0\) (SESAM calls it \(\kappa_c^{sea}\)) as
\begin{equation}
\frac{1}{2 \kappa_c^0} - 8 +
    \delta m \left(\alpha_s, \frac{1}{2\kappa_c^0}, \frac{1}{2\kappa_c^0} \right)
    = 0 \, .
\end{equation}
Then, suppressing the dependence on $\alpha_s$, 
\begin{eqnarray}
m a &=& \frac{1}{2\kappa^V} - \frac{1}{2\kappa_c^0} 
        + \delta m \left(\frac{1}{2\kappa^V},   \frac{1}{2\kappa^{sea}} \right)
        \nonumber \\
    &&  \hphantom{\frac{1}{2\kappa^V} - \frac{1}{2\kappa_c^0}}
        - \delta m \left(\frac{1}{2\kappa_c^0}, \frac{1}{2\kappa_c^0} \right) \, .
\end{eqnarray}
Near $a=0$, \(\kappa^V\) and \(\kappa^{sea}\)
approach \(\kappa_c^0\), $i.e.$ the valence and the sea quark
masses measured in lattice units approach zero, so we expand $\delta m$
about \(\kappa^V = \kappa^{sea} = \kappa_c^0\) 
\begin{eqnarray}
m a &=&        \left(\frac{1}{2 \kappa^V} - \frac{1}{2 \kappa_c^0}\right) + 
         \zeta^V \left(\frac{1}{2 \kappa^V} - \frac{1}{2 \kappa_c^0}\right) \nonumber \\
    && \qquad {} +
         \zeta^{sea} \left(\frac{1}{2 \kappa^{sea}} - \frac{1}{2 \kappa_c^0}\right) 
         \nonumber \\
    &\equiv& Z_m^{lat} \left( \frac{1}{2 \kappa^V} - \frac{1}{2 \kappa_c} \right),
    \label{eq:ma}
\end{eqnarray}
where 
\begin{eqnarray}
Z_m^{lat} & = & 1 + \zeta^V \nonumber\\
\frac{1}{2\kappa_c} & = & \frac{1}{2 \kappa_c^0} -
    \frac{\zeta^{sea}}{1+\zeta^V}
            \left(\frac{1}{2 \kappa^{sea}} - \frac{1}{2 \kappa_c^0}\right) .
\end{eqnarray}
Note that both $Z_m^{lat}$ and $\kappa_c$ are the partially quenched
ones.  What has happened is that, for Wilson like fermions, along with
the linear divergence which is absorbed in $1/2\kappa_c^0$, there is a
finite piece of $O(1)$ that shifts the quark mass. As this finite
piece is of $O(\mbar)$ after extrapolation to the physical sea quark
masses, its effect on the determination of $\mbar$ is dramatic, while
it is a small correction in $m_s$.

In the degenerate case Eq.~\ref{eq:ma} becomes 
\begin{eqnarray}
m a & = & ( 1 + \zeta^V + \zeta^{sea} ) \left(\frac{1}{2\kappa}-\frac{1}{2\kappa_c^0}\right)
          \nonumber\\
    & = & Z_m^{deg} \left(\frac{1}{2\kappa}-\frac{1}{2\kappa_c^0}\right) ,
\end{eqnarray}
where $Z_m^{deg} \equiv 1 + \zeta^V + \zeta^{sea} $ has a contribution
from valence and sea quarks.  At 1-loop, \(\zeta^{sea} = 0\),
therefore \(Z_m^{lat} = Z_m^{deg}\) and \(\kappa_c = \kappa_c^0\).
What the SESAM data are telling us is that corrections to this 1-loop
result are large, $i.e.$ \(\zeta^{sea} \sim 1 \).  Thus, we arrive at
the conclusion that combining the 1-loop $Z_m^{lat}$, which is
independent of $\zeta^{sea}$, with the partially quenched
extrapolation, which already incorporates $\zeta^{sea}$ as shown in
Eq.~\ref{eq:ma}, is more appropriate.  Another way of saying this is
that had $Z_m^{deg}$ been calculated non-perturbatively, one would
have found $Z_m^{deg} \sim 2 Z_m^{lat}$, which would have compensated
for the smaller $\mbar$ obtained from the degenerate extrapolation.

SESAM, by measuring quark masses with respect to $\kappa_c^0$, extract
$\mbar = 2.7(2)$ MeV and $m_s = 140(20)$. The large shift in
$m_s/\mbar$ from \CPT\ values is attributed to the effect of sea
quarks, which could change as $a \to 0$. Our proposal for using the
partially quenched extrapolation gives $\mbar = 4.7(1)$ MeV which is
$\sim 10\%$ below the quenched value at the same scale $a$ and roughly
preserves the \CPT\ ratio. Note, however, when using the partially
quenched method there exist enhanced chiral logs, similar to those in
the quenched approximation~\cite{PQCPTSharpe}. These afflict the
theory at small quark masses, but for present masses this effect
is expected to be negligible.

If the meson masses are completely independent of $\kappa^{sea}$, then 
\begin{equation}
{B_\pi^{sea}}/{B_\pi^{V}} = {B_\rho^{sea}}/{B_\rho^{V}} \,.
\end{equation}
The SESAM data bears this out within errors. However, when
this constraint is used in the fits for the vector masses, they
find high $\chi^2$s of about 50 for 25 degrees of freedom. We take
this statistically significant, but small, effect to indicate that
higher order terms in the chiral expansion do bring in small
dependence of the meson masses on the sea quark mass. 

In view of the above discussion, the only $n_f=2$ numbers that survive
the above discussed ambiguity from the 1996 analysis are for staggered
fermions, $\mbar \sim 2.7$ MeV.  More data are necessary to establish
the continuum limit in that case.  For Wilson like fermions the final
word on the values of $\mbar$ and $m_s$ has to await data using the WI
method along with a non-perturbative determination of the $Z$'s.

\section{Sum rule determinations of $\mbar$ and $m_s$}
\label{s:sumrules}

Progress in the sum-rules determination of quark masses has been
incremental as has been the case for LQCD. Over time the perturbation
expansion for the 2-point hadronic correction functions has been
carried out to higher order, along with a better determination of
\(\Lambda_{QCD}^{(3)}\). Models for the hadronic spectral functions
have been improved.  The main limitation continues to be the lack of
experimental data, with the one exception of $\tau$ decays.  Thus, one has to
model the spectral function rather than measure it. This slow but
steady progress was, as summarized in Table~\ref{t:sumrules}, reaching
a consensus by 1996 at $\mbar (2 \GeV) \approx 5 \MeV$ and $m_s (2 \GeV)
\approx 140 \MeV$.

Sum rule calculations proceed in one of two ways. (i) Using axial or
vector current Ward identities one writes a relation between two
2-point correlation functions, where the constant of proportionality
are the quark masses~\cite{BPR95,Jamin95}. (ii) Evaluate a given correlation function both
by saturating with known hadronic states and by evaluating it in
perturbative QCD (PQCD)~\cite{Narison95}. The PQCD expression depends on quark masses,
and defines the scheme in which they are measured.  Systematic errors
arise from the (i) finite order calculation of PQCD expressions, (ii)
the scale $\mu$ above which perturbative and hadronic solutions are
valid and can be matched on the average (duality), and (iii) the
ansatz for the hadronic spectral function.\looseness-1

\begin{table} 
\caption{Values and bounds on $\mbar$ and $m_s$ from sumrules.}
\setlength\tabcolsep{0.25cm}
\begin{tabular}{rcc}
\hline
reference             &$\mbar$ (MeV)    &$m_s$ (MeV)        \\
\hline
\cite{Narison89} 1989 &$ {}=   6.2(0.4)$&$ {}=   138(8)    $\\
\cite{BPR95}     1995 &$ {}=   4.7(1.0)$&$                 $\\
\cite{Narison95} 1995 &$ {}=   5.1(0.7)$&$ {}=   144(21)   $\\
\cite{Jamin95}   1995 &$               $&$ {}=   137(23)   $\\
\cite{Chetyrkin} 1996 &$               $&$ {}=   148(15)   $\\
\cite{Colangelo} 1997 &$               $&$ {}=   91 - 116  $\\
\cite{Jamin97}   1997 &$               $&$ {}=   115(22)   $\\
\cite{Prades97}  1997 &$ {}=  4.9(1.9) $&$                 $\\
\cite{Yndurain}  1997 &$ {}\geq 3.8-6  $&$ {}\geq 118 - 189$\\
\cite{Dosch97}   1997 &$ {}\geq 3.4    $&$ {}\geq 88(9)    $\\
\cite{Lellouch97} 1997&$ {}\geq 4.1-4.4$&$ {}\geq 104-116  $\\
\hline
\end{tabular}
\label{t:sumrules}
\vskip-0.75cm
\end{table}


In the last year, with the calculation of $\alpha_s^3$ terms in the
perturbative expansions, the value of \(\Lambda_{QCD}^{(3)}\) settling
around $380 \MeV$, and a critical reappraisal of the systematic errors
in the sum-rule calculations \cite{SR96BGM}, there has been a flurry
of activity as shown in Table~\ref{t:sumrules}.  The highlights of 
the new works are as follows. 

The calculation of $O(\alpha_s^3)$ terms and a detailed analysis of
the convergence of the perturbation expansion suggests that the
associated error is under control at $\approx 10\%$ level for $\mu \ge
2 $ GeV \cite{Chetyrkin}.

Colangelo et al.~\cite{Colangelo} have extended the analysis of $m_s$
in \cite{Jamin95,Chetyrkin} by constructing the hadronic spectral
function from known phase shift data. Similarly, Jamin~\cite{Jamin97}
has also used a different parametrization of the hadronic spectral
function using this phase shift data. In both cases the reanalysis
lowers the estimate of the strange quark mass significantly as shown
in Table~\ref{t:sumrules}.

Prades \cite{Prades97} has repeated the analysis of $\mbar$
incorporating the $\alpha_s^3$ corrections and using
\(\Lambda_{QCD}^{(3)} = 380 \) MeV. He reports a slightly higher value
than in \cite{BPR95}. This is because Prades chooses the duality point
at $\mu^2=2 \GeV^2$, where $\mbar$ has a maximum. There is a
significant decrease with $\mu$, the number dropping to $4.3(1.7) $ at
$\mu^2=3 \GeV^2$, and $3.8(1.5) $ at $\mu^2=4 \GeV^2$. The rationale
for the low choice of $\mu$ is that contributions not included in the
spectral function will bolster the answer for larger $\mu$. This
assumption needs to be substantiated.

Finally, a number of calculations have used the positivity of the
spectral function to derive lower bounds
\cite{SR96BGM,Yndurain,Dosch97,Lellouch97} which depend on $\mu$.  Of
these, the most stringent were reported by Lellouch at this conference
\cite{Lellouch97,Lat97Lellouch}. These bounds rule out the 1996
$n_f=2$ lattice results for $\mu \lsim 2.8$ GeV.  The hard to resolve
question is -- what is the scale $\mu$ at which PQCD, and thus the
bound, becomes reliable?  Unfortunately, this question cannot be
answered at present. \looseness-1

\section{Bottom quark mass}

There are two determinations of $m_b$. The NRQCD collaboration
\cite{Mb94NRQCD} determine it from the Upsilon binding energy and the
APE collaboration \cite{Mb97HQET} use HQET.
The two results agree:
\begin{eqnarray}
  m_b^{\MSbar}(m_b^{\MSbar}) &=& 4.15(5)(20)\ \GeV \quad{\rm (APE)} \nonumber\\
                             &=& 4.16(15)\ \GeV    \quad{\rm (NRQCD) } 
\end{eqnarray}

The nice features of the NRQCD method are (i) determination of the
scale from the spin averaged $1P-1S$ splitting. Data show that these
splittings are independent of the precise tuning of the input mass
$m^0_H$, and of the light quark action. (ii) tuning of $m^0_H$ using the
kinetic mass
\begin{equation}
   M = \lim_{p \to 0} \left({\partial^2 E}/
                                 {\partial p^2}\right)^{-1}.
\label{eq:Mkin}
\end{equation}
The heavy quark mass is then defined in two ways
\begin{eqnarray}
  m_H^{pole} &=& \frac{1}{2}[m_{H\bar H} - (E_{sim} - 2 E_0)] \nonumber\\ 
  m_H^{\MSbar} &=& Z^{pole\to \MSbar} m_H^{pole} \nonumber \\
  m_H^{\MSbar} &=& Z_m m_H^0 \,.
\end{eqnarray}
The drawbacks are that $Z^{pole\to \MSbar}$, $Z_m$, and $E_0$ (the
energy of a zero-momentum quark state) are only known in PQCD to 1-loop. The
two methods give consistent results, however, it is essential that the
calculation be done at other values of $\beta$, to test for stability
under variations of $a$.

The APE collaboration uses HQET to define
\begin{equation}
m_b(m_b) = \left(M_B {- {\cal E}} + {\alpha_s(a) X\over a}\right)
     \left(1 - {\alpha_s(m_b)\over \pi}\right). \nonumber
\end{equation}
They explain how the linear divergence in the energy measured on the
lattice, ${\cal E}$, is cancelled by that in the perturbative series
whose first term is $\alpha_s(a) X /a$. The residual renormalon
ambiguity is cancelled by a similar one in the second factor that
relates the pole mass to the $\MSbar$ mass.  However, to take into
account effects of higher orders in perturbation theory, they assign a
systematic error of $200 \MeV$ to the final result.  We believe that
their analysis of the cancellation of the renormalon ambiguity also
applies to the NRQCD analysis, and with a similar residual
uncertainty. The last issue concerning APE data is the stability
under variations of the lattice spacing. We feel that by averaging
over data at $\beta=6.0-6.4$, which shows a marginally significant
variation, the APE collaboration may have missed an equally important
source of uncertainty.

\section{Charm quark mass}

There are two new determinations of the charm quark mass by the APE
\cite{Mq97Giusti} and the Fermilab
collaborations~\cite{Mc97Kronfeld}. The extraction by APE uses the
same method as in \cite{Allton94}. The new features are that they use
both the WI and the HS methods and use non-perturbative estimates of
$Z$'s in the WI method. The drawback once again is that they average
the data at $\beta=6.0, 6.2$, and are not able to resolve
discretization errors.

The Fermilab collaboration uses a combination of the NRQCD and
Fermilab approaches. In their TI clover results, they include $O(ma)$ corrections in the
determination of 1-loop $Z$'s, and calculate the $q^*$ in the matching
factor using the Brodsky-Lepage-Mackenzie prescription, which they
argue gets rid off possible infrared scales seen in the connection to
pole mass. They find agreement in the extrapolated value between the
two ways of calculating $m_c$, and also argue that quenching errors
are expected to be small, unlike in the case of light quarks.

The final results are 
\begin{eqnarray}
m_c^{\MSbar}(m_c^{\MSbar}) &=& 1.525(40)(125) \GeV \quad {\rm APE} \ .
   \nonumber\\
m_c^{\MSbar}(m_c^{\MSbar}) &=& 1.33(8) \GeV \qquad {\rm Fermilab} \ .
\end{eqnarray}
We do not consider the difference significant as it could 
easily be due to the different ways of setting the scale and/or  
due to the discretization errors that the APE data does not 
resolve. 

Finally, we would like to mention the work of Bochkarev and Forcrand
\cite{Mc96Forcrand}. They calculate $m_c$ by evaluating the
correlation functions arising in QCD sum-rules on the lattice.  Their
estimate, in the quenched approximation, is
\begin{equation}
m_c^{\MSbar}(m_c^{\MSbar}) = 1.22(5) \GeV \ .
\end{equation}
The errors include effects of discretization and finite volume, and in
fact are dominated by the scale uncertainty. Given that these results
were obtained on relatively small lattices ($16^3 \times 32$) and with
low statistics (20 configurations), the quoted accuracy is
impressive.  This approach should be investigated further. 

\section{Conclusions and Acknowledgements}

Reliable estimates of quark masses have two immediate phenomenological
consequences. (i) Standard Model predictions of $\epsilon'/\epsilon$
are very sensitive to $m_s + m_d$ \cite{CP97Buras}, and (ii)
they constrain Supersymmetric models \cite{MSSM96Raby}. It is therefore 
exciting to report that the range listed in the Particle Data 
Book \cite{PDB96} has been significantly reduced. 

In the last year the quenched estimates have been significantly
improved, mainly due to the factor of 100 better data by the CP-PACS
collaboration.  There has been progress by the APE and ALPHA
collaborations in the non-perturbative determination of
renormalization constants.  The SESAM analysis has lead to a deeper
understanding of how to extract quark masses from $N_f=2$ simulations,
however, we are no further along in obtaining continuum limit
estimates.  Since the quenching errors are the least well understood,
it is this issue that needs maximum attention.

There has been progress in understanding errors in heavy quark analysis 
by the APE, Fermilab, and NRQCD collaborations.  We now have ``first generation'' 
estimates of $m_c$ and $m_b$. 

An area that has seen no progress is the study of isospin breaking
effects.  Knowing whether $m_u = 0$ is important since a zero value
solves the strong CP problem.  The exploratory quenched calculations
by Duncan {\it et al.}\ \cite{Mq96Duncan}, where an electromagnetic
field was added to quantify isospin breaking, reported a non-zero
value of $m_u$: $(m_d - m_u)/m_s = 0.0249(3)$ and $m_u/m_d =
0.512(6)$.  Unfortunately, quenched calculations cannot address this
question.  The subtle point is that quark masses calculated from low
energy phenomenology (or LQCD calculations with chiral extrapolations
based on truncated \CPT\ expansions) include instanton induced finite
renormalizations~\cite{Mq94Banks}. What one wants is $m_u$ defined at
a high scale.  LQCD can directly probe this, but only if simulations
are done including very light dynamical fermions whereby the influence
of instanton zero modes on the light quark propagation is properly
included. \looseness-1


Finally, it is exciting to see the rivalry between the LQCD and
sum-rules estimates of quark masses heating up.  We feel that both
sides have made big strides in understanding and addressing the
various sources of systematic errors, and estimates of masses have
been tightened.

We thank C.~Allton, L.~Giusti, S.~Gottlieb, S.~Hashimoto, H.~Hoeber, R.~Kenway,
A.~Kronfeld, M.~Luscher, S.~Ohta, K.~Schilling, G.~Schierholz, and
T.~Yoshie, for providing details of their data in advance. We also
acknowledge informative conversations with L.~Lellouch, M.~L\"uscher
and S.~Sharpe. We thank DOE and the ACL for support of our work.

\end{document}

%% file: rgmac.latex
\newskip\defaultbaselineskip\defaultbaselineskip=12pt

\def\GeV{\mathord{\rm \;GeV}}
\def\MeV{\mathord{\rm \;MeV}}

\def\bar{\overline}

\def\gsim{\mathrel{\raise2pt\hbox to 8pt{\raise -5pt\hbox{$\sim$}\hss{$>$}}}}
\def\rsim{\mathrel{\raise2pt\hbox to 8pt{\raise -5pt\hbox{$\sim$}\hss{$>$}}}}
\def\lsim{\mathrel{\raise2pt\hbox to 8pt{\raise -5pt\hbox{$\sim$}\hss{$<$}}}}
\def\ssqr#1#2{{\vbox{\hrule height.#2pt
      \hbox{\vrule width.#2pt height#1pt \kern#1pt\vrule width.#2pt}
      \hrule height.#2pt}\kern-.#2pt}}

\def\Dsl{\,\raise.15ex\hbox{$/$}\mkern-13.5mu D} 
\def\dsl{\raise.15ex\hbox{$/$}\kern-.57em\hbox{$\partial$}}

\ifx\href\undefined\def\href#1#2{{#2}}\fi
\def\spireshome{http://www.slac.stanford.edu/cgi-bin/spiface/find/hep/www?FORMAT=WWW&}
{\catcode`\%=12
\xdef\spiresjournal#1#2#3{\noexpand\protect\noexpand\href{\spireshome
                          rawcmd=find+journal+#1%2C+#2%2C+#3}}
\xdef\spireseprint#1#2{\noexpand\protect\noexpand\href{\spireshome rawcmd=find+eprint+#1%2F#2}}
\xdef\spiresreport#1{\noexpand\protect\noexpand\href{\spireshome rawcmd=find+rept+#1}}
\xdef\spireskey#1{\noexpand\protect\noexpand\href{\spireshome key=#1}}
}
\def\eprint#1#2{\spireseprint{#1}{#2}{#1/#2}}

\def\nohref{}
\def\onlyhref#1{}

\def\putpaper{\edef\refpage{\the\count0}%
              \def\nohref{}\def\onlyhref##1{}%
              {\def\ {+}\def\nohref##1{}\def\onlyhref##1{##1}%
               \edef\temp{\noexpand\spiresjournal
               {\journalname}{\volume}{\refpage}}\expandafter}\temp
               {\sfcode`\.=1000{\journalname} {\bf \volume} (\refyear)
                \refpage}\egroup}
\def\putpage{\edef\refpage{\the\count0}%
              \def\nohref{}\def\onlyhref##1{}%
              {\def\ {+}\def\nohref##1{}\def\onlyhref##1{##1}%
               \edef\temp{\noexpand\spiresjournal
               {\journalname}{\volume}{\refpage}}\expandafter}\temp
              {\refpage}\egroup}
\def\dojournal#1#2 (#3){\def\journalname{#1}\def\volume{#2}\def\refyear
                        {#3}\afterassignment\putpaper\bgroup\count0=}
\def\morepage{\afterassignment\putpage\bgroup\count0=}

\def\NPB#1{\dojournal{Nucl.\ Phys.}{B#1}}
\def\NPBPS#1{\dojournal{Nucl.\ Phys.\ \nohref{(PS)}\onlyhref{Proc.\ Suppl.}}
             {\nohref B#1}}

\def\PRL#1{\dojournal{Phys.\ Rev.\ Lett.}{#1}}
\def\PR#1{\dojournal{Phys.\ Rev.}{#1}}
\def\PRep#1{\dojournal{Phys.\ Rep.}{#1}}

\def\PRD#1{\dojournal{Phys.\ Rev.}{D#1}}

\def\PL#1{\dojournal{Phys.\ Lett.}{#1B}}

\def\PLB#1{\dojournal{Phys.\ Lett.}{#1B}}

\def\ZPC#1{\dojournal{Z.\ Phys.}{C#1}}

\def\etal{{\it et al.}}



%% file: t-ape.tex
\def\q{\quad}
\def\s{\phantom{-}}
\newcommand\0{hphantom{0}}
\newcommand\Zrat{$Z_A/Z_P$}
\begin{tabular}{|l|c|c|c|}
\hline
              & $\beta=6.0$   & $\beta=6.2$   &  $\beta=6.4$ \\
\hline
$Z_m$( pert)  &  1.252    &  1.222   &  1.200    \\
\Zrat( pert)  &  1.178    &  1.156   &  1.140    \\
\Zrat(Npert)  &  1.659    &  1.507   &  1.345    \\
$m_s$(WI) (MeV) &  128(6)   &  117(8)  &  107(8)   \\
$m_s$(HS) (MeV) &  133(16)  &  130(16) &  120(16)  \\
\hline
$Z_m$( pert)  & 1.177     &  1.157  &   1.143  \\
\Zrat( pert)  & 1.825     &  1.680  &   1.611  \\
\Zrat(Npert)  & 2.350     &  2.033  &   1.700  \\
$m_s$(WI) (MeV) & 125(5)    &  127(9) &   106(8) \\
$m_s$(HS) (MeV) & 117(7)    &  120(9) &   103(9) \\
\hline
\end{tabular}